\documentclass[aps,preprint]{revtex4}
\usepackage{graphicx}
\usepackage{dcolumn}
\usepackage{amsmath}
\usepackage{amssymb}

\def\ec{\epsilon_c}
\def\ex{\epsilon_x}
\def\exc{\epsilon_{xc}}
\def\be{\begin{equation}}
\def\ee{\end{equation}}
\def\bea{\begin{eqnarray}}
\def\eea{\end{eqnarray}}

\begin{document}
\title{Two-dimensional electron gas: correlation energy versus
density and spin polarization}
\author{Paola Gori-Giorgi, Claudio Attaccalite, Saverio Moroni, and
Giovanni B. Bachelet}
\affiliation{INFM Center for
  Statistical Mechanics and Complexity and
Dipartimento di Fisica, Universit\`a di Roma ``La Sapienza'', 
Piazzale A. Moro 2, 00185 Rome, Italy}
\date{\today}
\begin{abstract}
We propose a simple analytic representation of the correlation energy
$\ec$ for the two-dimensional electron gas, as a function of the density
parameter $r_s$ and the spin polarization $\zeta$. 
This new parametrization includes most of the known high- and low-
density limits and fits our new fixed-node diffusion Monte Carlo simulations,
performed for a wide range of electron densities ($1\le r_s \le 40$)
and spin-polarization states ($0\le \zeta\le 1$).
In this way we provide a reliable local-spin-density energy functional for
two-dimensional systems. The corresponding correlation potential is discussed
and compared with previous models.
\end{abstract}
\pacs{boh}
\maketitle
\section{Introduction}
The ideal two-dimensional electron gas (2DEG) is a simple model
in which $N$ strictly 2D electrons are confined in a square of surface
$S$ (periodically repeated in space) and interact via a $1/r$ potential
within a uniform, rigid neutralizing background. 
When studying this model, one is usually interested in its macroscopic
properties, i.e., the thermodynamic limit ($N,S\to \infty$ keeping $n=N/S$
constant) of its extensive physical quantities per particle.
Two parameters are enough to define the zero-temperature phase diagram of
the 2D electron gas, namely the density parameter
$r_s=1/{\sqrt{\pi n}a_B}$ (where $n$ is the density and $a_B$ the Bohr
radius) and the spin polarization $\zeta=(n_{\uparrow}-
n_{\downarrow})/n$, where $n_{\uparrow(\downarrow)}$ is the
density of spin-up (down) electrons.

The model itself is interesting, since it can provide information about
electrons confined in two dimensions realized in semiconductor 
heterostructures~\cite{ando}. 
Moreover, just like the three-dimensional case, the correlation energy
of the 2D electron gas as a function of density $r_s$ and spin polarization
$\zeta$ provides the local-spin-density (LSD) energy functional for 
density functional calculations of 2D systems. Currently, 2D LSD functionals
are based on parametrized diffusion Quantum Monte Carlo (DMC) data~\cite{TC}
at $\zeta=0$ and $\zeta=1$~\cite{usanoTCePZlike,usanoTC,DFT2D,TCperB0}. 
At intermediate spin polarizations, $0<\zeta<1$,
an exchange-like interpolation is often used~\cite{usanoTCePZlike}. 

We have recently presented new DMC simulations for a wide 
range of electron densities $r_s$ and spin polarizations $\zeta$~\cite{noi}.
The direct DMC calculation of the $\zeta$ dependence is new and provides
a reliable basis for building an LSD energy functional for 2D systems.
In this work we present and discuss an accurate parametrization of these new
data as a function of $r_s$ {\em and} $\zeta$. This new parametrization
accurately reproduces the $\zeta$ dependence of the DMC data and includes most
of the known high- and low-density limits. 
We also compare the corresponding correlation
potential to previous approximations, finding significant discrepancies
at $\zeta \neq 0$.

Hartree atomic units are used throughout this work.

\section{Diffusion Monte Carlo data}
Our calculations use standard fixed--node diffusion Monte Carlo
(FN-DMC)~\cite{mitas}, which projects the lowest-energy eigenstate
$\Phi$ of the many-body Hamiltonian with the boundary condition that
$\Phi$ vanishes at the nodes of a trial function $\Psi$.  Details of
the simulation are similar to those of Ref.~\cite{varsano}; further
details can be found in Ref.~\cite{noi}.
For each of the densities corresponding to $r_s=1, 2, 5, 10$ we have
considered about 20 values of $N$ and 10--12 polarizations
$\zeta$. For the densities $r_s=20$ and 30 we have used the
data of Ref.~\cite{varsano}. We have also computed the energy
at $r_s=40$ for $\zeta=1$.
To estimate the difference $\Delta$
between the energy $\epsilon_N(r_s,\zeta)$ of the finite system and
its thermodynamic limit $\epsilon(r_s,\zeta)$ we adopted a new strategy. 
Rather than a separate size extrapolation for each density
based on variational energies~\cite{TC,rapisarda,kwon}, we performed
a global fit directly based on FN-DMC energies, which exploits two
physically motivated ingredients: (i) the Fermi-liquid-like size
correction~\cite{ceperley78}
\be 
\Delta(r_s,\zeta,N) =  \epsilon_N(r_s,\zeta) -
\epsilon(r_s,\zeta) = 
\delta(1 + \lambda\zeta^2) 
[t_N(r_s,\zeta)-t_s(r_s,\zeta)]-{(\eta+\gamma\zeta^2)}/{N}
\label{delta}
\end{equation}
($t_N$ and $t_s$ are the Fermi energies of $N$ and $\infty$
particles, respectively, and $\delta, \lambda, \eta, \gamma$ are
$r_s$-dependent parameters); (ii) an
analytic expression for $\epsilon(r_s,\zeta)$, detailed in the next section,
which involves 12 more free parameters.  

The only uncontrolled source of error, the 
fixed--node approximation, depends on the nodal structure of $\Psi$. 
We choose a  Slater-Jastrow trial function with plane waves
(PW) as single orbitals. However, within the fixed-node approximation,
better results are obtained with backflow (BF) correlations in the wave
function~\cite{kwon}. Since simulations with
the BF wave function are considerably more demanding than with PW
determinants, we calculated BF energies only for $\zeta\!=\!0,
N\!=\!58$ and $\zeta\!=\!1, N\!=\!57$ for each density.
For other values of
$N$ and $\zeta$ the effect of backflow is estimated as a quadratic
interpolation in $\zeta$ and appended to PW energies, under the
further assumption that the size dependence be the same for BF and 
PW~\cite{noi}.


\section{Analytic model for the correlation energy}
In this section we present our parametrization of the correlation
energy of the 2D gas as a function of $r_s$ and $\zeta$.  
We first discuss the $\zeta$ dependence at a given fixed density,
then our choice for the $r_s$ dependence is presented, and finally
we impose the exact high- and low-density limits to our functional
form.
\subsection{Spin-polarization dependence}
We first noticed that, for $r_s\gtrsim 5$, the $\zeta$ dependence of the
exchange-correlation energy $\exc=\epsilon-t_s$ of our DMC
data is accurately described by a biquadratic form,
$c_0(r_s)+c_1(r_s)\zeta^2+c_2(r_s)\zeta^4$ (see also Ref.~\cite{varsano}).  
On the other hand,
the known high-density limit~\cite{mike}, 
\be 
\exc(r_s\to 0,\zeta)  =  \ex(r_s,\zeta)
 + a_0(\zeta)+b_0(\zeta)r_s\ln r_s+O(r_s),
\label{expansion}
\end{equation}
contains non-negligible contributions from higher powers of $\zeta$:
the dominating exchange term $\epsilon_x$ goes like
$(1+\zeta)^{3/2}+(1-\zeta)^{3/2}$, and the constant
term $a_0(\zeta)$ is well fitted by an eighth-degree
polynomial function of $\zeta$~\cite{mike}.  Since we want to interpolate the
energy between high and low density, we choose a functional form which
quenches the contributions to $\epsilon_x$ beyond fourth order in
$\zeta$ as $r_s$ increases,
\be
\ec(r_s\zeta)  =  \left(e^{-\beta r_s}-1\right)\,\ex^{(6)}(r_s,\zeta)
 + \alpha_0(r_s)+\alpha_1(r_s)\zeta^2+\alpha_2(r_s)\zeta^4,
\label{eq_exczdep}
\end{equation}
where
$$\ex^{(6)}(r_s,\zeta)=\ex(r_s,\zeta)-(1+\tfrac{3}{8}\zeta^2+
\tfrac{3}{128}\zeta^4)\ex(r_s,0)$$
is the Taylor expansion of $\epsilon_x$ beyond fourth order in
$\zeta$.  Since the first term in the righ-hand-side of
of Eq.~(\ref{eq_exczdep}) contains power 6 and higher of $\zeta$, 
it is
immediate to identify the function $\alpha_0(r_s)$ as the correlation
energy at zero polarization,
$$\alpha_0(r_s)=\epsilon_c(r_s,0).$$ 
Furthermore, $$2\,\alpha_1(r_s)=
\frac{\partial^2}{\partial\zeta^2}\ec(r_s,\zeta)\Big|_{\zeta=0}$$
gives the spin stiffness, and $$24\,\alpha_2(r_s)=
\frac{\partial^4}{\partial\zeta^4}\ec(r_s,\zeta)\Big|_{\zeta=0}.$$
\subsection{Density dependence}
We have now to fix the $r_s$ dependence of the functions $\alpha_i$. 
We generalize
the Perdew and Wang~\cite{PW92} form 
(designed for the three-dimensional gas) to the 2D case as follows
\be
\alpha_i(r_s) = A_i+(B_ir_s+C_ir_s^2+D_ir_s^3)\ln\left(1+\frac{1}{E_ir_s
+F_ir_s^{3/2}+G_ir_s^2+H_ir_s^3}\right).
\label{eq_alpha}
\end{equation}
This form possesses the small- and large-$r_s$ expansions:
\bea
\alpha_i(r_s\to 0) & = & A_i-B_i\,r_s\,\ln r_s + O(r_s)\\
\alpha_i(r_s \to \infty) & = & A_i+\frac{D_i}{H_i}+\left(\frac{C_i}{H_i}
-\frac{D_iG_i}{H_i^2}\right)\frac{1}{r_s}-\frac{D_iF_i}{H_i^2}\frac{1}
{r_s^{3/2}}+O\left(\frac{1}{r_s^2}\right),
\end{eqnarray}
and it thus has the correct 
high- and low-density behavior~\cite{mike,bonsall}, provided that the
constraint $A_i+D_i/H_i=0$ is imposed.
\subsection{Exact Limits}
Our $\ec(r_s,\zeta)$ has the correct functional form for small
and large $r_s$; it is now straightforward to impose most of the known
quantitative constraints.
We constrain our $\ec(r_s,\zeta)$ to fulfill: 
(i) the requirement that the exact values~\cite{HD,mike} of 
$a_0(\zeta)$ and $b_0(\zeta)$ at
$\zeta=0$ and $\zeta =1$ in the small-$r_s$ expansion of
Eq.~(\ref{expansion}) are recovered, which implies
\bea
A_0 & = & -0.1925 \\
B_0 & = & \frac{\sqrt{2}}{3\pi}(10-3\pi) \\
A_0+A_1+A_2+a_x\beta{\cal F}(1) & = & -0.039075 \\
B_0+B_1+B_2 & = & \frac{10-3\pi}{12\pi},
\end{eqnarray}
where
\bea
{\cal F}(\zeta) & = & (1+\zeta)^{3/2}+(1-\zeta)^{3/2}-
\left(2+\tfrac{3}{4}\zeta^2
+\tfrac{3}{64}\zeta^4\right) \label{eq_F}\\
a_x & = & \frac{4}{3\pi\sqrt{2}}; \label{eq_ax}
\end{eqnarray}
 (ii) the requirement that the total energy $\epsilon(r_s,\zeta)$ be
independent of $\zeta$ for $r_s \to \infty$
up to order $O(r_s^{-2})$, thus recovering the low-density
behavior $\epsilon \to - m/r_s + n/r_s^{3/2} + O
(r_s^{-2})$~\cite{bonsall} with positive $m$ and $n$ independent of
$\zeta$. We thus have
\bea
A_i+\frac{D_i}{H_i}  =  0 \\
\frac{C_1}{H_1}-\frac{D_1G_1}{H_1^2}  =  \frac{3}{4}a_x \\
\frac{C_2}{H_2}-\frac{D_2G_2}{H_2^2}  =  \frac{3}{64}a_x \\
F_1 = F_2 = 0.
\end{eqnarray}
We also fixed $A_1$ according to the high-density limit of the 
spin susceptibility~\cite{HD,mike}, and $G_2=0$ since it turned out 
to be an irrelevant parameter in our fitting procedure.
In this way, we have built an analytic model which interpolates between
the exact high- and low-density limits and has 12 free parameters to be
fixed by a best fit to our diffusion Monte Carlo data.
We then perform a global fit $(r_s,\zeta,N)$, which also includes the 
infinite size extrapolation of Eq.~(\ref{delta}), to our data set (122 data for
$1\le r_s \le 40$, $0\le \zeta \le 1$, and $21\le N \le 114$). In this
way we fix the values of 36 free parameters, 24 of which disappear from
the final analytic expression of $\ec$ since they only concern the $N\to
\infty$ extrapolation. This fit yields a reduced $\chi^2$ of 3.8. The
optimal values of the parameters which yield the model for $\ec$ of
the infinite system are reported in Table~\ref{tab_ecinf}.

\section{LSD correlation potential}
\begin{figure}[h]
\includegraphics[width=\columnwidth]{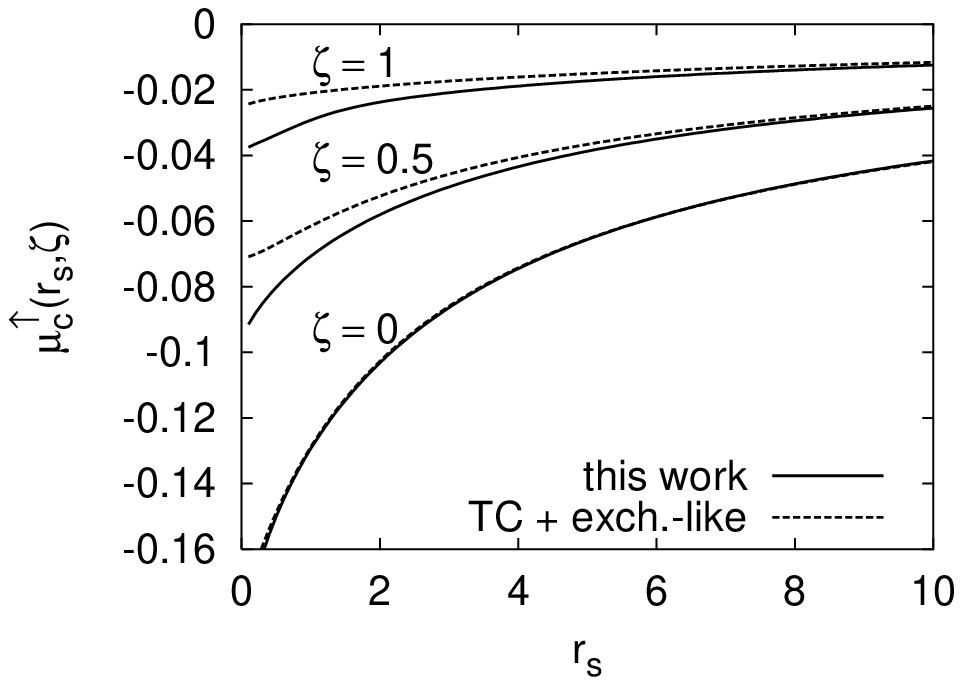} 
\caption{}
\label{fig_corrpot}
\end{figure}

The 2D LSD correlation potential $\mu_c^{\sigma}$
for electrons of spin $\sigma$ is given by
\be
\mu_c^{\sigma}(r_s,\zeta)=\frac{\partial [n\ec(r_s,\zeta)]}
{\partial n_{\sigma}}=\ec(r_s,\zeta)-\frac{r_s}{2}\frac{\partial
\ec(r_s,\zeta)}{\partial r_s}-(\zeta-{\rm sgn}\,\sigma)
\frac{\partial\ec(r_s,\zeta)}{\partial \zeta},
\label{eq_pot}
\end{equation}
where ${\rm sgn}\,\sigma$ is $+1$ for spin-$\uparrow$ electrons
and $-1$ for spin-$\downarrow$ electrons. From our parametrization
of $\ec(r_s,\zeta)$ we get:
\be
\frac{\partial \ec(r_s,\zeta)}{\partial r_s}=
a_x{\cal F}(\zeta)\frac{\left[e^{-\beta r_s}(1+\beta r_s)-1\right]}
{r_s^2}+\alpha_0'(r_s)+\alpha_1'(r_s)\zeta^2+\alpha_2'(r_s)\zeta^4,
\end{equation}
where ${\cal F}(\zeta)$ and the constant $a_x$ are given by Eqs.~(\ref{eq_F}) 
and~(\ref{eq_ax}), respectively, and
\bea
\alpha_i'(r_s) & = & \frac{d\alpha_i}{dr_s}=
(B_i+2C_ir_s+3D_ir_s^2)\ln\left[1+\frac{1}{f_i(r_s)}\right]
-\frac{(B_i+C_i r_s^2+D_i r_s^3) f_i'(r_s)}{f_i(r_s)[f_i(r_s)+1]} \\
f_i(r_s) & = & E_ir_s+F_ir_s^{3/2}+G_ir_s^2+H_ir_s^3 \\
f_i'(r_s) & = & E_i+\tfrac{3}{2}F_i r_s^{1/2}+2G_ir_s+3H_ir_s^2.
\end{eqnarray}
The derivative w.r.t. $\zeta$ is simply
\bea
\frac{\partial \ec(r_s,\zeta)}{\partial \zeta} & = & \frac{a_x}{r_s}\left(1-e^{
-\beta r_s}\right){\cal F}'(\zeta)+2\alpha_1(r_s)\zeta+4\alpha_2(r_s)\zeta^3
\\
{\cal F}'(\zeta) & = & \tfrac{3}{2}\left(\sqrt{1+\zeta}-\sqrt{1-\zeta}\right)
-\tfrac{3}{2}\zeta-\tfrac{3}{16}\zeta^3.
\end{eqnarray}
It is interesting to compare our correlation potential with the approximations
used in previous LSD calculations in two 
dimensions. The most used 2D LSD functional is the one given
by Tanatar and Ceperley~\cite{TC}, who performed diffusion Monte Carlo
simulations at $\zeta=0$ and $\zeta=1$, and gave an analytic fit
of the corresponding correlation energies. For the $\zeta$ dependence, many
authors~\cite{usanoTCePZlike} used the exchange-like approximation
\be
\ec(r_s,\zeta)=\ec(r_s,0)+\frac{\left[(1+\zeta)^{3/2}+(1-\zeta)^{3/2}-2
\right]}{2^{3/2}-2}[\ec(r_s,1)-\ec(r_s,0)].
\label{eq_exchlike}
\end{equation}
In Fig.~\ref{fig_corrpot} we compare our correlation potential (as a function
of $r_s$ and for three different values of the spin polarization $\zeta$)
with this widely used {\em Tanatar-Ceperley-plus-exchange-like}
correlation potential. One can see that, while for 
$\zeta=0$ the two potentials are
almost indistinguishable, for $\zeta \neq 0$ there are significant
discrepancies: at $\zeta=1$, the difference between the two potentials
is $\sim$~30\% at $r_s=1$; for lower densities this difference is lower, 
being 15\% at $r_s=4$ and 7\% at $r_s=10$. At
$\zeta=1$ the discrepancies do not in fact
depend on the exchange-like choice for the $\zeta$
dependence: they are exclusively
due to the corresponding correlation energy of 
Tanatar and Ceperley, which differs from ours between 35\% and 4\% for
$r_s\in [0,10]$. To test the intrinsic quality of the 
exchange-like interpolation
against our new $\zeta$-interpolation scheme, we plugged into
Eq.~(\ref{eq_exchlike}) our new correlation
energies at $\zeta=0$ and $\zeta=1$. As shown in Fig.~1 of Ref.~\cite{noi},
the $\zeta$ dependence of QMC data is rather different, especially at
lower densities: at the density of the transition to the fully polarized 
gas, $r_s\sim 26$, the exchange-like
interpolation predicts an energy barrier between the $\zeta=0$ and 
the $\zeta=1$ phases which is more than an order of magnitude higher
than the QMC result.

Our correlation energy which at $\zeta=1$ is, as said, quite
different from the Tanatar-Ceperley~\cite{TC} value,
should be much closer to the
true one since
(i) we included the effect of backflow on the nodes, (ii) we
imposed the exact high-density limit, and (iii) the infinite size
extrapolation is directly performed on a DMC data set.

\section{Summary and conclusions}
We have presented a new, reliable, LSD functional for 2D systems,
based on a new set of DMC data for a wide range of electron
densities and spin polarizations, and on an analytic form which
efficiently reproduces these data and includes most of the known
high- and low-density limits. A
comparison of the corresponding correlation potential with previous
approximations show, for $\zeta\neq 0$, differences up to 30\% for
$r_s \in [0,10]$.

\section*{Acknowledgements} 
We acknowledge partial financial support from MURST (the Italian 
Ministry for University, Research and Technology) through COFIN99.

\section*{captions}
\begin{itemize}
\item{Table I - 
Optimal fit parameters for the correlation
energy, as parametrized
in Eqs.~(\ref{eq_exczdep}) and~(\ref{eq_alpha}). Values
labelled with $^*$ are obtained from exact conditions.
The parameter $D_i=-A_iH_i$ is not listed (see text).}
\item{Figure 1 - Correlation potential for spin-up electrons as a function
of the density parameter $r_s$ and for three different values of the 
spin polarization $\zeta$. The present result is compared with
the exchange-like interpolation~\cite{usanoTCePZlike} applied
to the Tanatar and Ceperley data~\cite{TC}.}
\end{itemize}

\newpage


\begin{table}[h]
\begin{tabular}{lccc}
\hline\hline
 &  $i=0$ & $i=1$ & $i=2$  \\
\hline
 $A_i$ &$-0.1925^*$    & $0.117331^*$              &$0.0234188^*$   \\  
 $B_i$ & $0.0863136^*$ &$-3.394 \times 10^{-2} $    &$-0.037093^*$  \\  
 $C_i$ & $0.0572384$    &$-7.66765\times 10^{-3*}$             &$0.0163618^*$\\ 
 $E_i$ & $1.0022 $      &  $0.4133 $                  &$1.424301$      \\  
 $F_i$ & $-0.02069$    & $0^*$                     & $0^*$       \\  
 $G_i$ & $0.33997$       & $6.68467\times10^{-2}$       & $0^*$    \\  
 $H_i$ & $1.747\times 10^{-2}$      & $7.799\times 10^{-4}$  & $1.163099$   	 \\
 \\
 $\beta$ & $1.3386$ &  \\ 

\hline\hline  
\end{tabular}
\caption{}
\label{tab_ecinf}
\end{table}

\newpage

\end{document}